\documentclass{camera}

\usepackage{graphicx}
\usepackage{amsmath}
\usepackage{amssymb}

\begin{document}

%

\title{Magnetic monopole searches in the cosmic radiation}

%
\author{Ivan De Mitri}

\organization{Dipartimento di Fisica - Universit\`a di Lecce \\
and Istituto Nazionale di Fisica Nucleare - Sezione di Lecce, Italy}

\maketitle

%

\abstract{There has been a big effort in the past twenty years with at least a couple of 
generations of experiments which searched for supermassive GUT magnetic monopoles in the 
cosmic radiation. Here a short review of these searches 
is given, together with a brief description of the theoretical framework and of the 
detection techniques.}

\section{Introduction and theoretical framework}

While the existence of magnetic monopoles is not excluded by classical electromagnetism,
the first convincing argument in favor of such particles was made by Dirac in 1931,
showing that the existence of one single monopole could explain the 
observed quantization of the electric charge \cite{monorev}.
A big jump in the magnetic monopole research was made in the seventies when 't Hooft 
and Polyakov \cite{monorev} showed that each time a 
semi-simple non abelian gauge group 
({\it e.g.} SU(N)) is broken leaving a residual U$_{em}$(1) subgroup, magnetic monopoles 
are produced as topologically stable soliton solutions of the theory. 
Their mass $m$ is of the order of the energy scale at which the symmetry breaking takes 
place. 
A cosmological production of magnetically charged point-like topological
defects (via the Kibble mechanism) is then foreseen in the framework of 
Grand Unified Theories (GUT). This results in a flux of super-massive monopoles 
with $m \sim 10^{17} \,$GeV. Lower masses ($m \sim 10^{10}\div 10^{15} \,$GeV) 
can results from GUT theories with intermediate 
scale symmetry breaking \cite{monorev}.
At our time monopoles can be searched for in the penetrating cosmic radiation as
``fossil" remnants of these early transition(s).

Unfortunately no definite prediction can be made on the monopole flux.
It would be either too large in classical cosmology (the so called {\it monopole problem}),
or too low, practically undetectable, in an inflationary scenario, or at a 
measurable level if thermal production is assumed after the inflationary phase.
However, some upper limits can be obtained from arguments based on 
magnetic field survival or mass density.
By requiring that monopoles do not short-circuit the galactic magnetic field faster 
than the dynamo mechanism can regenerate it, an upper limit on their flux can be obtained. 
This is the so-called Parker bound ($\sim 10^{-15} \,$cm$^{-2}$s$^{-1}$sr$^{-1}$ 
\cite{monorev}), whose value sets the scale of the detector exposure for  monopole 
searches. Under some assumptions, an Extended Parker Bound (EPB) at the level of 
$1.2\times 10^{-16}(m/10^{17}\,$GeV) cm$^{-2}$s$^{-1}$sr$^{-1}$ was also 
obtained \cite{monorev}.
Another upper limit can be obtained by requiring that the monopole flux is such that
the related mass density do not overclose the Universe (the {\it closure bound}).
If, as  suggested by Rubakov and Callan \cite{monorev}, GUT magnetic monopoles 
catalyze nucleon decays along their path with a cross section
$\sigma_c $ of the order of the hadronic cross sections, a different set of flux limits
can be obtained based on the observed luminosity of astrophysical objects 
({\it e.g.} neutron stars) in which monopoles could be gravitationally captured.
These limits are very stringent (few orders of magnitude below the Parker bound) but 
they rely on strong assumptions on the physical properties (e.g. magnetic field strength 
and configuration) of the interested region \cite{monorev}.

The velocity range in which GUT magnetic monopoles should be sought spreads over 
several decades.
If sufficiently heavy ($m \gtrsim 10^{16} \,$GeV), they
would be gravitationally bound to the galaxy with a velocity distribution
peaked at $\beta~=~v/c~\simeq 10^{-3}$. 
Lighter  monopoles, with masses around $10^{7}\div 10^{15} \,$GeV,
would be accelerated to relativistic velocities in one or more coherent 
domains of the galactic magnetic field, or in the intergalactic field, or 
in several astrophysical sites like a neutron star \cite{monorev}.

\section{Detection techniques and experimental searches}
Experiments aiming to perform a sensitive GUT monopole search below the Parker bound
need very large acceptances ({\it i.e.} thousands of m$^2$sr), good sensitivity in a wide velocity 
range going from $\beta \simeq 10^{-5}$ up to $\beta \simeq 1$, and livetimes of the order of 
at least few years. 
These needs forced the optimization of the detection methods which can be classified into
{\it Induction}, {\it Energy Loss} and  {\it Catalysis} based {\it Techniques}
(hereafter IT, ELT and CT respectively). These have been used for both direct
and indirect searches for monopoles. Here we will briefly go through their main results.

A magnetic charge passing through a loop of wire induces a current jump that can be 
subsequently detected. This detection principle is by far the the best one for the search for 
magnetic monopoles since it does not depend on other characteristics like velocity, mass, 
electric charge, and it works also if the monopole comes with an attached proton or heavier nucleus
(these bound systems might form due to the magnetic charge-moment interaction).
Nevertheless the difficulty and cost of building large arrays limited this technique to 
sensitivities to fluxes orders of magnitude above the Parker bound.
The combination of the IT results gives an upper limit
\footnote{All the upper limits will be given at $90 \%$ C.L.}
 of $\sim 2 \cdot 10^{-13}$ cm$^{-2}$s$^{-1}$sr$^{-1}$ \cite{monorev}.

The energy loss of GUT magnetic monopoles across different kind of materials strongly depends 
on their velocity \cite{monorev}.
Therefore different detectors and analysis strategies have to be adopted to search for monopoles
in different $\beta$ ranges. The peculiar property of a fast magnetic monopole 
with $\beta \gtrsim 10^{-2}$ is its large ionization power compared either to the 
considerably slower monopoles or to minimum ionizing electrically charged particles.
On the other hand, slow magnetic monopoles should leave small signals spread over a
large time window. It is therefore difficult to have good sensitivity in a wide velocity 
range within a single experiment. Many experiments have been done (or are currently running)
in order to perform direct or indirect searches for monopoles by using ELT's in various
$\beta$ regions.

\begin{table}
  \begin{center}
     \begin{tabular}{ccc}\hline
{\bf Experiment}        &  {\bf  Detection Technique}   &  {\bf Location}    \\ \hline
Kolar Gold Field      &    gas detectors  & KGF mine (India)  \\ 
Baksan         & scint. counters & Baksan valley (Russia)  \\ 
Soudan         &    gas detectors  &  Minnesota (USA) \\ 
Ohya           &  CR-39 track-etch  &  Ohya quarry (Japan) \\ 
MACRO          &  scint. + gas + track-etch & Gran Sasso (Italy) \\ 
AMANDA         & {\u C}erenkov light & South Pole \\ 
Baikal         & {\u C}erenkov light & Baikal lake (Russia) \\ \hline
     \end{tabular}
     \caption{\label{tab:directELT}
              List of the main experiments that are (or have been) performing direct searches 
              with {\it Energy Loss Techniques} \cite{monorev}.}          
  \end{center}
\end{table}

\begin{table*}[ht]
  \begin{center}
    \begin{tabular}{ccccc}
      \hline
      {\bf Experiment} & {\bf Technique} & {\bf Flux limit} & {\bf $\beta$ range} \\
      \hline
      Soudan1   & gas detectors &
      $8.8\cdot 10^{-14}$ & $10^{-2} \div 1$ \\
      IMB  & {\u C}erenkov light &
      $1\div 3\cdot 10^{-15} $ & $10^{-5} \div 10^{-1}$ \\
      Kamiokande  &  {\u C}erenkov light &
      $2.5\cdot 10^{-15}$ & $5\cdot 10^{-5} \div 10^{-3}$ \\
      Baikal  &  {\u C}erenkov light &
      $6\cdot 10^{-17}$ & $\sim 10^{-5}$ \\
      MACRO  & Streamer tubes &
      $3\cdot 10^{-16}$ & $1.1\cdot 10^{-4} \div 5\cdot 10^{-3}$ \\
      \hline
    \end{tabular}
    \caption{\label{tab:catalimits}
             Flux upper limits limits (in $cm^{-2} s^{-1} sr^{-1}$) obtained by
             experiments searching for nucleon decays catalyzed by a GUT monopole
             (reprinted from \cite{macrocat}).}
  \end{center}
\end{table*}

Direct searches have been performed by a number of experiments, each of them exploiting a 
different aspect of the ELT. Some of them are listed in tab.\ref{tab:directELT}, together
with the detection technique used in the search. Here we will very briefly describe two of 
them, namely MACRO and AMANDA.

\begin{figure}[h]
\vskip -2.4cm
\begin{center}
  \includegraphics[width=3.5in,angle=0]{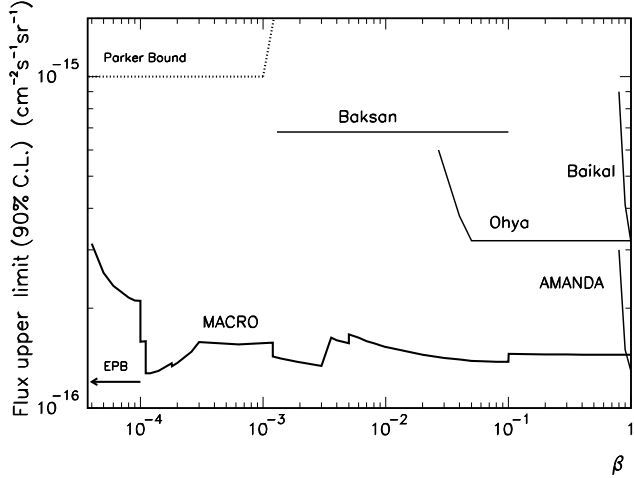}
\end{center}
\vskip -0.5cm
\caption{\label{fig:limiti}
          The global 90\% C.L. upper limits to an isotropic flux 
          of bare magnetic monopoles obtained by several experiments
          in direct searches, without the assumption of monopole induced nucleon
          decay catalysis (reprinted from \protect\cite{macro}).}
\end{figure}

MACRO was a large multipurpose underground detector mainly optimized for the search for GUT
magnetic monopoles with velocity $ \beta \geq 4\times 10^{-5} $ and with a
sensitivity well below the Parker bound.
The apparatus, which took data up to December 2000, was arranged in a modular structure and had 
a total acceptance of $\sim 10,000 \,$m$^2$sr.
Redundancy and complementarity in monopole searches were provided by the use of three independent
detection techniques: scintillation counters, limited streamer tubes and track-etch detectors.
Dedicated hardware and analysis procedures were adopted to search for monopoles in different
$\beta$ ranges with different subdetectors.
This allowed, for the first time, a sensitivity well below the Parker bound in a very wide 
velocity range. As shown in fig.\ref{fig:limiti}, a limit of 
$\sim 1.5 \cdot 10^{-16}$ cm$^{-2}$s$^{-1}$sr$^{-1}$ was put for monopoles with 
$\beta \gtrsim 10^{-4}$ (see \cite{macro} for further details).

AMANDA is an array of strings of photomultipliers, located deep under the ice of South Pole,
mainly devoted to the search for very high energy neutrinos. 
It is sensitive to monopoles through the detection of the huge amount of {\u C}erenkov 
light emitted by relativistic ($\beta \gtrsim 0.8$) magnetic charges \cite{amanda}.
Due to the high transparency of the ice and to the dimensions of the array, the acceptance
is very large, then a good sensitivity to very low fluxes is ensured.
However, due to the detection principle itself, such sensitivity is confined in a very narrow 
region around $\beta \sim 1$ \cite{amanda}.
This is nonetheless interesting because of the possible connection between relativistic light
GUT monopoles and cosmic ray events above the GZK cutoff (see sec.\ref{sec:discuss}).

Indirect searches can be performed by looking for monopole induced tracks in ancient
mica crystals. If a monopole-nucleus bound system crosses a piece of such materials a permanent 
damage is produced along the trajectory, that can be subsequently evidenced through a
chemical etching. Since these crystals have been exposed for very long times ($\sim 10^8 \,$yr), 
very strong upper limits, at the level of $10^{-18}$ cm$^{-2}$s$^{-1}$sr$^{-1}$,
can be obtained \cite{price}.
However, since the detection threshold is large, this technique is sensitive in a limited $\beta$ 
region around $\beta \sim 10^{-3}$ and the formation of a stable monopole-nucleus bound system
must be assumed. 
In other kinds of indirect searches, the experiments look for monopoles trapped in bulk matter 
({\it e.g.} iron sand) \cite{monorev}.

If GUT monopoles catalyze nucleon decays along their path  with a sufficiently 
large cross section, direct searches can be performed also by means of the detection of the decay 
products (CT). As a consequence many of the detectors originally built for nucleon decay searches
have been used to look for monopoles. Detectors like AMANDA or Baikal, are also sensitive to
monopoles by detecting the  {\u C}erenkov light emitted by the decay products.
MACRO performed a sensitive search for catalysing magnetic monopoles with its streamer tube
system \cite{macrocat}. The results of these searches are  summarized in tab.\ref{tab:catalimits}.

\section{Discussions and conclusions}
\label{sec:discuss}
The null result obtained by several experiments in a wide velocity range put very stringent
limits to the local monopole flux that will be hardly improved by other searches and that
represent a strong weir against future theoretical speculations. 
Some of the more stringent results, referring to direct searches 
for bare monopoles, are shown in fig.\ref{fig:limiti}. As can be seen
the MACRO result covers the whole region ensuring the highest sensitivity and putting
a strong upper limit, while underwater/ice 
experiments seem very promising in the ultra-relativistic regime only 
(unless nucleon decay catalysis is assumed).
Strong upper limits have also been put with indirect searches or under the hypothesis of a 
monopole induced catalysis of nucleon decays.

Since theory is unable to give a reasonable prediction on the expected flux,
with a negative result one cannot put stringent conclusions on GUT and/or cosmology, except
for some superstring models that foresee fluxes at the level of the Parker bound \cite{monorev}.
A renewed interest in the field has been triggered by some models that would explain
the anomalous flux of Ultra High Energy Cosmic Rays (UHECR) observed above the GZK cutoff as 
due to magnetic monopoles.
There are essentially two classes of models. In the first scenario UHECR would have been produced 
in the annihilation of monopole-antimonopole pairs. The binding mechanism
results however to be very inefficient, unless poles are connected by strings, and in this case
they would not be detectable with standard techniques since they would not show a 
magnetic charge \cite{blanco}.
In the second case, light ultrarelativistic monopoles would produce showers in crossing 
the atmosphere thus simulating a UHECR event. A monopole flux at the level of the Parker 
bound would be sufficient to explain the observed rates \cite{wick}.
If this is the case important progress might be done in the next few years by underwater/ice 
experiments, due to their large acceptances and sensitivities in the relativistic region.

\vskip 0.2cm
\noindent {\bf Acknowledgments} \\
The author gratefully acknowledges the MACRO Collaboration, in particular the
members of the {\it rare particle working group}.


%
\end{document}